\documentclass[12pt]{article}
\usepackage{epic}
\usepackage{rotating}
\usepackage{subfigure}
\usepackage{psfig}
\usepackage{epsfig}
\renewcommand{\d}{\delta}
\newcommand{\ctc}{$|c_2|^2$-curve}
\newcommand{\btc}{$G_2$-curve}
\newcommand{\pbtc}{processed $G_2$-curve}

\newcommand{\ket}[1]{|#1\rangle}
\newcommand{\ketpsi}[1]{|\psi(#1)\rangle}
\newcommand{\bra}[1]{\langle #1|}
\newcommand{\pro}[1]{\ket{#1}\bra{#1}}
\newcommand{\ex}[1]{\langle #1\rangle}
\title{A sequence of unsharp measurements enabling a real time
visualization of a quantum oscillation}
\author{J\"urgen Audretsch\thanks{E-mail: 
Juergen.Audretsch@uni-konstanz.de},   
Thomas Konrad\thanks{E-mail: Thomas.Konrad@uni-konstanz.de}, 
Artur Scherer\thanks{E-mail: Artur.Scherer@uni-konstanz.de}\\
\normalsize \it Fachbereich Physik der Universit\"at Konstanz\\
\normalsize \it Postfach M 673, D-78457 Konstanz, Germany\\
\normalsize 31.07.2000}
\date{}
\begin{document}
\maketitle
\begin{abstract}
The normalized state
$\ketpsi{t}=c_1(t)\ket{1}+c_2(t)\ket{2}$ of a single two-level system
performs oscillations under the influence of a resonant driving
field. It is assumed that only one realization of this process is
available. We show that it is possible to approximately visualize in
real time the evolution of the system as far as it is given by
$|c_2(t)|^2$.
For this purpose we use a sequence of particular unsharp measurements
separated in time. They are specified within the theory of generalized 
measurements in which observables are represented by positive operator
valued measures (POVM). A realization of the unsharp measurements may
be obtained by coupling the two-level system to a meter and performing
the usual projection measurements on the meter only.
\end{abstract}
\section{Introduction}
Recently, considerable progress has been made in the experimental and
theoretical study of single quantum systems in atomic traps and  optical
cavities as well
as in the context of solid-state physics (coupled quantum dots). 
In many cases the physical situation may be described by means of a two level system with states
$\ket{1}$ and $\ket{2}$ 
under the influence of a periodic driving potential $V(t)$. 
The resulting motion of the normalized state 
\begin{equation}
\label{psi}
\ket{\psi}=c_1\ket{1}+c_2\ket{2}
\end{equation}
involves oscillations of the probabilities $|c_1(t)|^2$ and
$|c_2(t)|^2$ called Rabi oscillations. The usual way to measure the
dynamics of $|c_2(t)|^2$ employs many projection measurements
of an observable with eigenstates $\ket{1}$ and $\ket{2}$. For this
purpose an ensemble
is prepared in the initial state and a projection measurement is
carried out at time $t_0$ on each member of the ensemble leading to
the determination of $|c_2(t_0)|^2$. Repeating the procedure for different
times $t$ one obtains the Rabi oscillations of $|c_2(t)|^2$.

This procedure fails if there is only one single quantum system 
available and the objective is to visualize its motion 
in real time. Imagine, e.g., there is a single two-level system
performing only once a hundred Rabi oscillations. Is there a method to 
visualize them? There are two complementary conditions such a method must meet. 
The measurement must not disturb the system too much, so that its
evolution remains close to the undisturbed case. On the other hand,
the coupling of the measurement apparatus to the two level system 
should nevertheless be so strong that the readout shows accurately
enough the
modified dynamics of the system (including the disturbance by the measurement). 

This can clearly not be achieved by standard measurements which
project onto an eigenstate of the measured observable. They disturb
the system too strongly. For instance, the well known
Quantum Zeno effect shows that it is possible to  detect perfectly the
modified dynamics of a system at the price of loosing all information about
the undisturbed motion. The measurement disturbs the dynamics so
much as to make the system stay in one of the eigenstates of the
observable. A more promising strategy to fulfill our
demands is to carry out appropriate sequences of generalized
measurements. Generalized measurements differ from projection
measurements in that the probabilities of their outcomes are in
general given by the expectation values of positive operators instead
of projection operators.
 
The theory of generalized measurements in which observables are
represented by positive operator valued measures (POVM) has
intensively been studied \cite{2}. For a survey 
see Busch et al.~\cite{3}. These measurements are in their generic
form also called unsharp, as we will do, or non-ideal, weak, soft or
fuzzy, thereby paying attention to one of their different properties. 
They have been investigated theoretically in connection  with the
problem of joint non-ideal measurements of incompatible
observables, c.f., Busch et al.~\cite{3}. Recent quantum optical
measurements like the Haroche-Ramsey set-up have been analyzed too with
respect to the question of complementarity  by de
Muynck and Hendrikx \cite{4}
using the concept of generalized measurements.
Martens and de Muynck \cite{1} address the question of which measurements extract
the most information about the 
system. 
Information extraction and disturbance has been widely discussed in
quantum information theory, as e.g., in the context of continuous
measurement and feedback control by Doherty et. al.~\cite{5}.
We will deal instead with sequences of discrete unsharp measurements which are 
separated in time and show that, if appropriately chosen, they may provide a real time
visualization of Rabi oscillations.

In the context of continuous quantum measurements of energy the 
question of visualization of oscillations has been addressed  in
\cite{6} in a phenomenological approach, 
which was based on the application of 
restricted path integrals \cite{7}.
In \cite{6} it had been indicated, that for continuous measurements 
a correlation may exist between the time-dependent measurement readout
and modified Rabi oscillations between energy eigenstates. This
visualization of a state evolution has been numerically verified in
\cite{8}, to our best knowledge for the first time (compare also \cite{12}).
Oscillations in coupled quantum dots measured by a
quantum point contact are treated in an approach to
continuous quantum measurements using stochastic master equations 
by Korotkov \cite{9}. For a quantum trajectory approach see Goan et
al.~\cite{11}. 

A realization
scheme for the continuous fuzzy measurement of energy and the
monitoring of a quantum transition has then been given in \cite{10}.
The intention was to present a microphysical
basis for the phenomenologically motivated continuous measurement
scheme. Inspired by this article we give below an independent
treatment which is solely based on successive single-shot measurements of the
generalized type (POVM). All concepts necessary to introduce an
appropriate measurement readout, to justify this choice and to specify
different measurement regimes are based on these single
measurements. This is also the case for the numerical evaluation which
makes use of the simulation of single unsharp measurements and the
otherwise undisturbed dynamical evolution (Rabi oscillations) between
the measurements. Our intention is to
keep our treatment as general as possible but nevertheless readable
for those who may be able to design experimental realizations. 
We will give an application
to a particular quantum optical setup in a subsequent paper
\cite{Audretsch00b}.

This paper is organized as follows: In Sect.~\ref{2} we specify the
particular subclass of unsharp measurements on which
our considerations are based. In Sect.~\ref{3} a succession of these  
unsharp measurements is studied. In addition we introduce
the concept  of a 'best guess' based on the outcome of one series of $N$
consecutive measurements (N-series). In Sect.~\ref{4} a Hamiltonian is
introduced generating the Rabi oscillations we want to measure by
means of series of unsharp measurements with time $\tau$ between two
consecutive 
measurements. An appropriate measurement readout is defined and
different regimes of measurement are specified. Finally the results of
the numerical analysis (specified in App.~C) are presented in
Sect.~\ref{5}. In order not to interrupt the flux of the presentation we
postponed the detailed motivation of the particular choice of
generalized measurements to App.~A and some technical
considerations concerning the measurement readout to App.~B.

\section{Single unsharp measurements}
\label{2}
For later use we introduce in this section a well known special class of
generalized measurements  and discuss how they may be experimentally
realized. 
 
\subsection{A particular class of single unsharp measurements}
\label{2.1}
We discuss measurements on a {\em two-level system} (S) with orthonormed states
$\ket{1}$ and $\ket{2}$, which are eigenstates of an observable  $A$
represented by a self-adjoint operator. This observable may be for
example energy or spin.  We perform a particular kind of generalized measurement, that is a
measurement which can be represented  by means of a
{\em Positive Operator Valued Measure} (POVM). Its operational meaning
will become clearer in Subsection \ref{2.2}, where we describe its
realization. The measurement is 
specified as follows:
There are only two {\em outcomes} or pointer readings denoted 
by $+$ and $-$. The duration of the measurement is $\d \tau$.
Depending on the outcome the normalized initial state
$\ketpsi{0}$ of S is tranformed due to  the measurement into
one of the related states $\ket{\psi_{\pm}(\d\tau)}$,
\begin{equation}
\label{statetrafo}
\ketpsi{0} \rightarrow \ket{\psi_{\pm}(\d\tau)} = M_{\pm}\ketpsi{0} 
\end{equation} 
with so-called  {\em operations}, which are required to be diagonal: 
\begin{eqnarray}
\label{defM+}
M_+ &:=& u_1^+\pro{1}+u_2^+\pro{2}\,, \\
\label{defM-}
M_- &:=& u_1^-\pro{1}+u_2^-\pro{2} \,.
\end{eqnarray}

We add to this first requirement an important second one:
$u_{1,2}^+$ and
$u_{1,2}^-$ are assumed to be positive. Here and in the following an overall phase factor $\exp(\mbox{i}\varphi_\pm)$ is
omitted without loss of
generality. A justification for both
requirements is given in App.~A. 
Let us  introduce 
\begin{equation}
\label{p1p2}
p_1:=(u_1^+)^2 =1-(u_1^-)^2\,, \quad p_2:=(u_2^+)^2=1-(u_2^-)^2\,.
\end{equation}
We will characterize the measurement later on by the {\em parameters}
\begin{equation}
\label{p0deltap}
p_0:= \frac{1}{2}(p_1 +p_2)\,,\quad\Delta p:= p_2-p_1\,.
\end{equation}

According to the specifications above, we restrict ourselves to a
particular class of {\em generalized measurements} with commuting hermitean
operations, which will prove to be useful for 
our purpose. 
Throughout the paper we discuss selective measurements, which are
non-destructive. We use the Schr\"odinger picture  unless otherwise
stated.  Note, that the state vectors will in general not be
normalized. 

The probabilities $p_\pm$ for the outcomes $+$ and $-$
to occur if the measurement is performed on the state $\ketpsi{0}$
are given by the expectation value of the so-called {\em effects}
\begin{eqnarray}
\label{effects1}
E_{\pm}&=& M^{\dagger}_{\pm}M_{\pm}\,,\\
\label{effects2}
E_+ +E_- &=& 1\,
\end{eqnarray}
according to     
\begin{equation}
\label{prob}
p_{\pm}= \ex{E_{\pm}}_{\psi(0)}\,.
\end{equation}
This leads to $0\le p_{1,2}\le 1$ and $p_++p_-=1$.
One may also write
\begin{eqnarray}
\label{defE+}
E_+&:=& p_1\pro{1}+p_2\pro{2}\,,\\
\label{defE-}
E_-&:=& (1-p_1)\pro{1}+(1-p_2)\pro{2}\,.
\end{eqnarray}
The effects $E_+$ and $E_-$ are semi-positive operators. Because of
the property (\ref{effects2}) they constitute a POVM. It is a
characteristic trait of the type of measurements in question that the
effects are in general not projections: $E^2_{\pm}\ne E_{\pm}$. We call
this the {\em genuine
case}. 

Only in the limiting cases
with $p_1= 1$ and $ p_2=0$ (or with indices $1$ and $2$
interchanged), i.e. $|\Delta p| =1$, we obtain a
projection operator valued measure (PVM). Looking at the resulting
states $\ket{\psi_{\pm}(\delta \tau)}$ in this limit we see that
they are identical with   
$\ket{1}$ or $\ket{2}$. Accordingly, if the initial state $\ket{\psi(0)}$ is
$\ket{2}$, we obtain reliably $+$  as measurement outcome and
$-$ for $\ket{1}$ (or with indices $1$ and $2$ interchanged). This well
known von Neumann measurement is called a 
{\em sharp measurement}.

The opposite limit is characterized by  $|\Delta p|\ll 1$. Then
$E_{\pm}$ are nearly proportional to the identity operator and the
probabilities $p_{\pm}$ of the outcomes 
become nearly independent of  the initial state $\ket{\psi(0)}$.
Even if the initial state is $\ket{1}$ (or $\ket{2}$),
there are finite
probabilities for both outcomes $+$ and $-$.   
Because in the genuine case the generalized measurement does not allow a fully
reliable conclusion regarding the initial state, it is called an
{\em unsharp (or non-ideal) measurement}. With $E_\pm$ of the form
(\ref{defE+}) and (\ref{defE-}) it is generally accepted to refer to it as an
unsharp measurement of the observable $A$ \cite{1}. Because it is not evident
what this notation could mean from an operational point of view, we
discuss an explicit realization below. In the limit $|\Delta p| \ll 1$, the
measurement changes the state only a little as can be seen from
equations (\ref{defM+}) and (\ref{defM-}). Regarding its influence on the state, an unsharp
measurement may also be called {\em weak} or {\em soft}. 
Below we will
make use of the fact that single measurements with different degrees of
weakness may be introduced by choosing the parameters $p_0$ and
$|\Delta p|$
appropriately.

\subsection{Realization of unsharp measurements}
\label{2.2}
It is known that the unsharp measurements above can be realized by
coupling the two-level systems S  to a {\em meter} M  and performing a projection
measurement on M only (c.f. Neumark's theorem \cite{3}).
It is to be expected, that in this way experiments with
unsharp measurements of the class we are discussing 
may in practice be performed for example in the domain of quantum
optics or condensed matter physics. We give therefore some details:

The two-level system in the state 
\begin{equation}
\label{psi0}
\ket{\psi(0)}=
c_1\ket{1}+c_2\ket{2}
\end{equation}
interacts with another two-level system with orthonormed
states $\ket{\Phi_\pm}$ which acts as the meter M. Its 
initial state is $\ket{\Phi(0)}$. Hence the initial state of the compound system is 
$|\Psi(0)\rangle = \ket{\psi(0)}\ket{\Phi(0)}$. It is then assumed that
the interaction lasting for the time $\d\tau$ results in the
particular unitary
development   
\begin{eqnarray}
\label{sysmeter}
|\Psi(0)\rangle &\rightarrow&  
\ket{\Psi(\delta \tau)}\\ &= &u_1^+c_1\ket{1}\ket{\Phi_+}+
u_1^-c_1\ket{1}\ket{\Phi_-} \nonumber\\&+& u_2^+c_2\ket{2}\ket{\Phi_+}
+u_2^-c_2\ket{2}\ket{\Phi_-}\nonumber 
\end{eqnarray}
with positive $u_{1,2}^+\,, u_{1,2}^-$ and $||\Psi(0)||=||\Psi(\delta \tau)||=1$.

After $\d\tau$ a sharp measurement on the meter alone is
performed which, depending on  the outcome $+$ or $-$, projects on the state $\ket{\Phi_+}$ or
$\ket{\Phi_-}$ with $P_\pm :=1\otimes \ket{\Phi_\pm}\bra{\Phi_\pm}$.
These two  pointer readings 
$+$ and $-$ are taken as the measurement outcomes of Section \ref{2.1}. The
readout $+$ occurs with probability 
\begin{equation}
\label{sharp}
p_+=\langle P_{+}\rangle_{\Psi(\delta \tau)} = (u_1^+)^2|c_1|^2 +(u_2^+)^2|c_2|^2\,,
\end{equation}
which agrees with $p_+$ of (\ref{prob}). In this case the state of the
compound system after the measurement is the product state
\begin{equation}
\label{finalstate}
\ket{\Psi_+(\d\tau)}= \ket{\Psi(\delta \tau)} =
(u_1^+c_1\ket{1}+u_2^+c_2\ket{2})\ket{\Phi_+} = \ket{\psi_+(\d\tau)}\ket{\Phi_+}
\end{equation} 
with $\ket{\psi_+(\d\tau)}$ as given in equations 
(\ref{statetrafo}) with (\ref{defM+}) and (\ref{defM-}).  In order to obtain the 
final state for the measurement readout $-$  one simply has to interchange
the indices $+$ and  $-$. 
The class of unsharp measurements of Section \ref{2.1} is thus
reconstructed.

The device specified above shows clearly that the unsharpness in
question is of genuine quantum nature.
It cannot be traced back to an imperfect measurement procedure, which
itself may for example be caused by
an imperfectly fixed pointer showing from time
to time $+$ although the meter state is $\ket{\Phi_-}$.
The reason is instead that the quantum dynamics leading to the unitary
development (\ref{sysmeter}) does not correlate $\ket{1}$ only with
$\ket{\Phi_+}$   and 
$\ket{2}$ only   with $\ket{\Phi_-}$  but allows the appearance of the states
$\ket{2}\ket{\Phi_+}$ and $\ket{1}\ket{\Phi_-}$ in the superposition.

\section{N-series of successive unsharp measurements}
\label{3}
\subsection{One N-series}
We will now perform on the same single system $N$ unsharp measurements
of the type specified above with the same parameters in an immediate
succession. This is called an {\em
N-series}. Later on we will allow the measurements to be separated in
time by $\tau$. A motivation to use N-series will be given below.
Since the operations $M_+$ and $M_-$ commute, the final state after
the N-series will not depend on the order of $+$ and $-$ results. For
any particular sequence of results with a total number $N_+$  of
results $+$ the state is transformed for $\tau=0$ according to 
\begin{equation}
\label{opnseries}
\ketpsi{0}\rightarrow M(N_+,N)\ketpsi{0}
\end{equation} 
with 
\begin{equation}
\label{opabrev}
M(N_+,N):= M_+^{N_+}M_-^{(N-N_+)}\,.
\end{equation} 

Now we discard the information about
the order of the results, i.e., we restrict ourselves to the
information that the total number of $+$ results is $N_+$, regardless 
when they occurred in the sequence. As measurement outcome attributed
to the total N-series we take in the following the relative frequency 
of positive results 
\begin{equation}
r:= \frac{N_+}{N}\,.
\end{equation}
In order to
transcribe for this type of 
unsharp measurement the scheme of Section \ref{2.1}, which is based on the
equations (\ref{statetrafo}), (\ref{effects1}) and (\ref{prob}) 
for non-normalized state vectors, we have to work out the related
operations and effects. The probability $p(N_+,N)$
that $N_+$ positive results are measured in an N-series is ${N
\choose{N_+}}$ times the probability that a particular ordered
sequence of $N_+$ positive and $N-N_+$ negative results is obtained:
\begin{equation}
\label{probN}
p(N_+,N) = \ex{E(N_+,N)}_{\psi(0)}\,,
\end{equation} 
where the effect $E(N_+,N)$ is 
given by 
\begin{eqnarray}
\label{efn1}
&&E(N_+,N) = 
         {N\choose{N_+}}\,M^{\dagger}(N_+,N)M(N_+,N) \\ 
         & = &
         {N\choose{N_+}}\,\left[p_1^{N_+}(1-p_1)^{(N-N_+)}\pro{1}
         +p_2^{N_+}(1-p_2)^{(N-N_+)}\pro{2}\right]\nonumber\,.
\end{eqnarray}
The corresponding probability is 
\begin{eqnarray}
\label{probn}
p(N_+,N)& = &{N\choose{N_+}}\left(p_1^{N_+}(1-p_1)^{(N-N_+)}|c_1|^2\right.
\\ &&  + \left. p_2^{N_+}(1-p_2)^{(N-N_+)}|c_2|^2\right)\,. \nonumber    
\end{eqnarray} 
To complete the scheme we write the pure
state resulting after the N-series in
the form 
\begin{equation}
|\psi_{N_+}\rangle = \sqrt{{N\choose{N_+}}}M(N_+,N)\ketpsi{0}\,. 
\end{equation}

In order to quantify by how much an N-series of measurements affects
the state, we calculate for arbitrary values of $p_1$ and $p_2$  the
fidelity, which shows how closely the post-measurement state resembles
the pre-measurement state. The fidelity $F(N)$ between the pure state 
$\ketpsi{0}$ before the measurements and the state 
\begin{equation}
\rho= \sum_{N_+=0}^N\,|\psi_{N_+}\rangle\langle \psi_{N_+}|
\end{equation}
after the N-series is equal to the square of root of the overlap
between $\ketpsi{0}$ and $\rho$ \cite{Nielsen.Chuang00}
\begin{equation}
F(N):= \sqrt{\langle \psi|\rho|\psi\rangle}\,.
\end{equation}
Please note, that $|\psi_{N_+}\rangle$ is unnormalized.
After some algebra one finds 
\begin{equation}
\label{fidelcon}
F(N)= \sqrt{1- 2|c_1|^2|c_2|^2 (1-b)}\,,
\end{equation}
where
\begin{equation}
b:= \left(\sqrt{p_1p_2} + \sqrt{(1-p_1)(1-p_2)}\right)^{N}\,.
\end{equation}
Equation (\ref{fidelcon}) shows that for all choices of the parameters
$p_1$ and $p_2$ the fidelity assumes its minimum for
$|c_1|^2=|c_2|^2=1/2$ and its maximum for $|c_1|^2=1$ and $|c_2|^2=0$
(or with indices interchanged). 

Fidelity may serve as a direct measure for the weakness of the
influence of an N-series of weak measurements on the pre-measurement
state. This becomes evident by looking at the limiting cases: for
extremely sharp measurements with $p_1=1$ and $p_2=0$ (or vice versa)
the fidelity equals the fidelity $F= \sqrt{1- 2|c_1|^2|c_2|^2}$
of a projection measurement. The maximal fidelity $F=1$ is obtained
for infinitely weak measurements with $p_1=p_2$.

For an infinite N-series the limit $\lim_{N\to\infty} F=\sqrt{1-
2|c_1|^2|c_2|^2}$ equals  for all values of $p_1=1$
and $p_2=0$ with $p_1\not=p_2$ the fidelity of an projection
measurement.  That such an iteration of measurements produces an
eigenstate has theoretically been discussed in
\cite{Sondermann98}. For an experimental scheme which realizes a
projection measurement by an iteration of unsharp measurements see
\cite{Brune.et.al90}.

\subsection{Best guess for the outcome of one N-series}
\label{3.2}
For the moment we refer again  to repeated measurements on the same
initial state $\ketpsi{0}$ (ensemble approach).
The statistical expectation value of $r=N_+/N$ reads
\begin{eqnarray}
\label{expn2}
{\cal E}(r) &=& \sum_{N_+=0}^N r(N_+) p(N_+,N)\\
             &=&  p_1|c_1|^2 + p_2|c_2|^2\,.\nonumber
\end{eqnarray}
The latter equation follows with (\ref{probn}). In a similar way we
obtain the variance $\sigma^2(r)$:

\begin{equation}
\label{varn}
\sigma^2(r) = |c_1|^2|c_2|^2(\Delta p)^2 + \frac{1}{N}\left(|c_1|^2
p_1(1-p_1)+|c_2|^2 p_2(1-p_2)\right)\,.
\end{equation} 
Based on (\ref{expn2}) it is easy to relate $|c_2|^2$ of the
initial state $\ketpsi{0}$ 
with the expectation value ${\cal E}(r)$
of an N-series of unsharp measurements starting with $\ketpsi{0}$:
\begin{equation}
\label{c2n}
|c_2|^2 = \frac{{\cal{E}}(r)-p_1}{\Delta p}\,.
\end{equation} 
Please note, that both quantities $|c_2|^2$ and ${\cal E}(r)$ refer to
an ensemble  represented by the state $\ketpsi{0}$.

If only one N-series is measured as will be the case below, only a {\em best
guess} $G_2$ for the quantity $|c_2|^2$ can be obtained. Equation (\ref{varn})
suggests to choose it as 
\begin{equation}
\label{b2}
G_2:= \frac{r-p_1}{\Delta p}\,.
\end{equation} 
The standard deviation of the best guess $G_2$ is then given by 

\begin{eqnarray}
\label{varb2}
\sigma(G_2) &=&\frac{\sigma(r)}{|\Delta p|}\\
            &=& \sqrt{|c_1|^2|c_2|^2 + \frac{1}{N}\frac{|c_1|^2
p_1(1-p_1)+|c_2|^2 p_2(1-p_2)}{(\Delta p)^2}}\,.\nonumber
\end{eqnarray}
It specifies how accurate $|c_2|^2$ of the pre-measurement
state $\ketpsi{0}$ can be estimated by means of one
N-series. $\sigma(G_2)$ can be decreased by increasing $N$ or
$|\Delta p|$.  

In case of a single measurement ($N=1$), the standard deviation
$\sigma(G_2)$ assumes its minimal value $|c_1||c_2|$ for $|\Delta p|
=1$, i.e., for a 
projection measurement ($p_1=1$ and $p_2=0$ or vice versa).
For all other choices of the parameters $p_1$ and $p_2$ ($|\Delta
p|<1$) and other choices of $N$, there is a larger standard deviation.
This justifies to call
these measurements unsharp, because they provide less reliable
information  about $|c_2|^2$. 

As we will see below, there are physical situations in which the 
parameters are fixed and have to be chosen with
$|\Delta p|<1$. Eq. (\ref{varb2}) shows that in this case the standard
deviation decreases with increasing number of repetitions $N$ so that it is favorable to
refer to N-series. This is the reason why we will now make use of them.

\section{Measurement of a dynamically driven state by a sequence of
N-series}
\label{4}
\subsection{Best guess as measurement readout}
We refer in the following not to an ensemble but to one single
two-level quantum system. It is driven by an external periodical
influence with frequency $\omega$ to
perform resonant Rabi oscillations, if it is not disturbed by measurements.
In the Schr\"odinger picture its undisturbed dynamics is therefore  given by 
\begin{equation}
\label{H}
H = a_1 \pro{1} +a_2\pro{2} + \frac{\hbar\Omega_R}{2}(\ket{2}\bra{1}\exp\{-i\omega t\}
+\ket{1}\bra{2}\exp\{i\omega t\})\, 
\end{equation} 
with $\Omega_R = 2\pi/T_R$ and Rabi period
$T_R$. Because of the resonance we have: $\omega = (a_2
-a_1)/\hbar$.

It is our intention to visualize the undisturbed behaviour in time of
$|c_2(t)|^2$, i.e., in our case the Rabi oscillations. 
A sharp measurement implies a 
drastic change of $\ketpsi{t_0}$ into $\ket{1}$ or $\ket{2}$. A
succession of projection measurements will in fact result
approximately in the Zeno effect. Unsharp measurements can be
weak and may change the
state less. If their parameters are properly adjusted, one can therefore expect
that these measurements are superior for our purpose. 
Accordingly many unsharp measurements of the type specified in
Sect.~\ref{2.1} are
performed on this system with time $\tau$ between two consecutive 
measurements. The unitary development between the measurements is determined
by the Hamiltonian $H$. It is assumed that the time $\tau$ is much larger than
the duration $\delta\tau$ of a single unsharp measurement ($\tau\gg
\delta \tau$). Again we bundle up $N$ unsharp measurements to an
N-series of total duration $\Delta t := N\tau$, so that in fact a sequence of
N-series is obtained with outcomes $r(t_m)$ at $t_m= m\Delta t$ with $m=
1\,,2\ldots$. In order to resolve the Rabi oscillations we require $\Delta t
\ll T_R$.  As outlined in Appendix B the best guess 
$G_2(t_m)$ related to $r(t_m)$ by 
\begin{equation}
\label{b2t}
G_2(t_m)= \frac{r(t_m)-p_1}{\Delta p}
\end{equation}
can still be used as approximation for $|c_2(t_m)|^2$ even if a
driving influence is present. However, it
will only be a good approximation if the number $N$  of measurements within
a N-series is not too large. More precisely, $N$ has to fulfill
inequality (\ref{Nbound}). The sequence
$\left(G_2(t_m)\right)$  obtained from the outcomes of the
successive N-series serves as the  {\em measurement readout}.

\subsection{Level resolution time and the different regimes of
measurement}
How unsharp (or weak) must the underlying single measurement of
Section \ref{2} for our purpose be?
A measure for the strength of the driving influence in (\ref{H}) is
given by the frequency $\Omega_R$. A strong influence corresponds to a
short Rabi period of the oscillations  
between the two levels $\ket{1}$ and $\ket{2}$. 
On the other hand measurements on the system, even if they are
unsharp, modify and hinder this primary time development of the state
of the system. We have therefore two contrary influences on the
two-level system. For a comparison we
want to introduce a characteristic time which represents the reciprocal
strength of the disturbing influence of the measurement alone. To this end we
discuss in a first step again one N-series with $\tau=0$.

The two states $\ket{1}$ and $\ket{2}$ of the two-level system
correspond to  $|c_2|^2=0$ and $|c_2|^2=1$. According to (\ref{varb2})
the standard deviation $\sigma(G_2)$ of the best guess $G_2$ improves
with increasing $N$. How large must $N$ be in order to  obtain a
standard deviation smaller than $1$, so that one can discriminate
between the two values of $|c_2|^2$ above? We have to demand 

\begin{equation}
\label{1.gt.sq1}
1 \geq \sigma^2 (G_2)\,.
\end{equation} 
$\sigma(G_2)$ depends on $|c_1|$ and $|c_2|$ of the initial
state. But this requirement should be fulfilled for any
initial state. We may therefore take in (\ref{1.gt.sq1}) the maximum
value of $\sigma(G_2)$ 
\begin{equation}
\label{1.gt.sq2}
1 \geq \frac{1}{4} + \frac{1}{N}\frac{
p_1(1-p_1)+p_2(1-p_2)}{2(\Delta p)^2}\,,
\end{equation}
which amounts to  
\begin{equation}
\label{1.gt.sq3}
N \geq \frac{4 p_0(1-p_0)}{3(\Delta p)^2} + \frac{4}{9}\,.
\end{equation}
In fact we are interested in unsharp measurements within the N-series
which are repeated
after a time $\tau>0$. In order to discuss the influence of the
measurements separately, we consider the absence of the driving
influence ($g=0$). Then the operator of the dynamical evolution
between the measurements commute with $M_\pm$. Therefore independent
from the initial state     
altogether a finite interval of time not larger than 
\begin{equation}
\label{Tlr}
T_{lr} = \tau N_{lr} =  \tau\frac{4 p_0(1-p_0)}{3(\Delta p)^2}  
\end{equation}
is needed in order to obtain a variance of the related best guess so small
as to resolve the levels with high probability. We call $T_{lr}$ the {\em level resolution time}.

We return to the situation when in addition the driving external
influence is active too. The complete setup is then characterized by two time
scales: the level resolution time $T_{lr}$ and the Rabi period $T_R$.
There are three physically different regimes: for  $T_{lr}\ll T_R$ the succession of
measurements is strong or sharp enough to push the state to one of the
states $\ket{1}$ or $\ket{2}$ before the Rabi oscillation becomes
manifest. Connected with this is a small variance. Therefore
$G_2(t_m)$ will correspond well to the disturbed  $|c_2(t_m)|^2$ but
the readout will not reflect the underlying Rabi oscillations. We call 
this the {\em quantum jump regime} for reasons to be seen later. 
For $T_R \ll T_{lr}$
on the other hand, the measurement influence on the state dynamics is
so weak, that the state and therefore
$|c_2(t_m)|^2$  shows the Rabi oscillations. But because of the large
variance of $G_2$, a single 
readout  $G_2(t_m)$ will in general not  be significantly related to 
$|c_2(t_m)|^2$ but will show a {\em fuzzy} behaviour.
The state motion thus cannot be read off from $G_2(t_m)$. We call this
the {\em Rabi regime}. Of particular interest is the {\em intermediate
regime} 
characterized by $T_{lr}/T_R\approx 1$. One can expect that
this regime shows two
features: on one hand the state motion is still close to the Rabi
oscillations and on the other hand the readout $G_2(t_m)$ follows
closely the $|c_2(t_m)|^2$ of the state. In this case the state motion
would be made visible as intended, without modifying too much the
original Rabi oscillations by the influence of the measurement.

The numerical simulations show that the
intermediate regime can rather be characterized by the condition 
\begin{equation}
\label{F}
f:=3\pi T_{lr}/T_R \approx 1
\end{equation}    
than by $T_{lr}/T_R\approx 1$. In the
following we use $f$ instead of $T_{lr}$ as {\em measure of fuzziness}.

\section{Results of the numerical analysis}
\label{5}
\begin{figure}
\centerline{
\epsfig{figure= 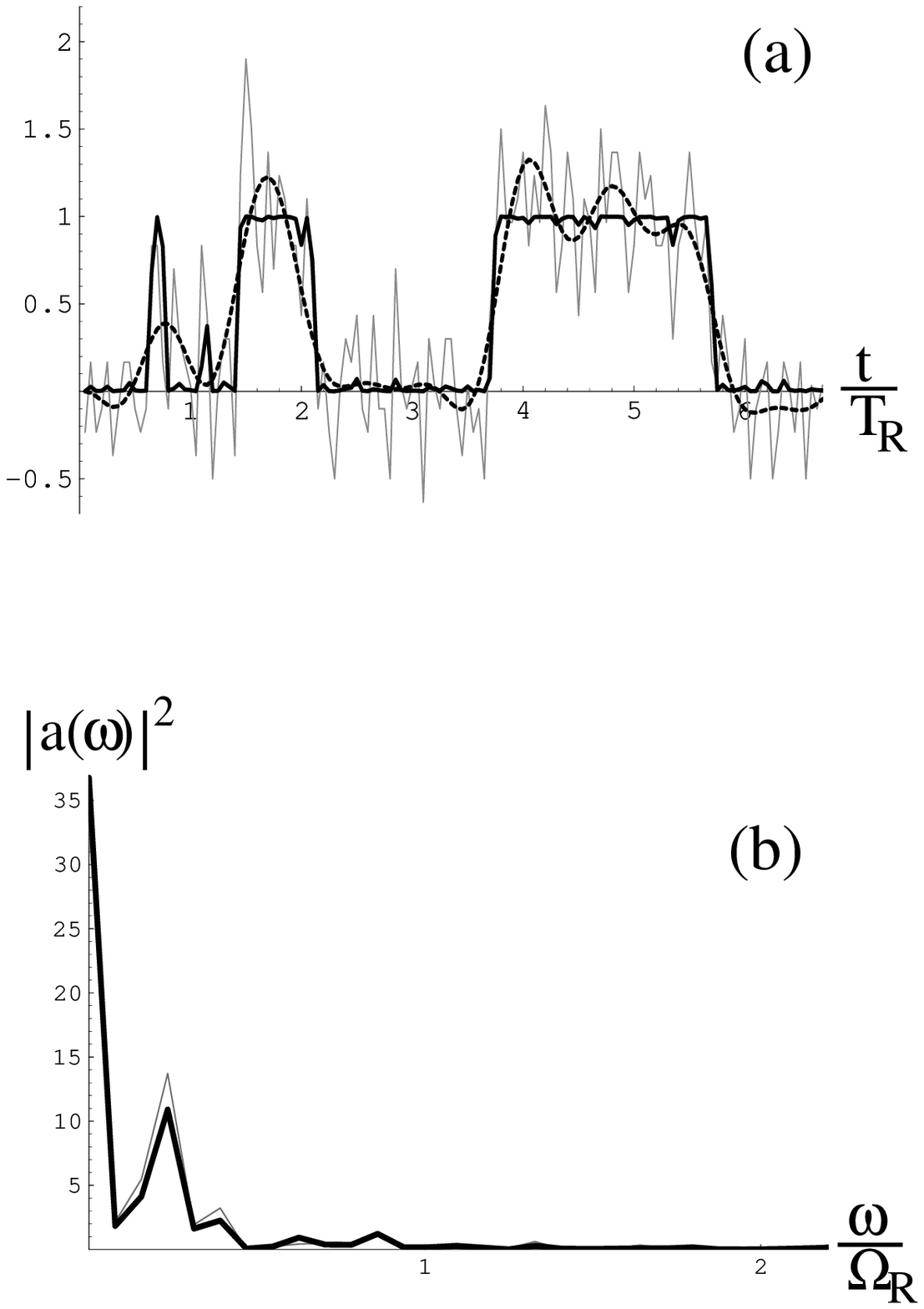, width=0.8\textwidth, angle=0}}
\caption{\label{figure1} Quantum jump regime with fuzziness
$f=0.07$ and parameters $\Delta p= -0.3\,, p_0=0.5\,, \tau= 0.002T_R\,,
N=25$. In Figure (a) black
represents the \ctc{}, gray the measurement readout $G_2(t)$ and the
dashed curve is obtained from the \btc{} by noise reduction. On the time
axis multiples of the Rabi period $T_R$ are indicated. Figure (b)
shows the power spectra of the \ctc{} (black) and the
measurement readout (gray). The
frequency $\omega$ is given in units of the Rabi period $\Omega_R$.  
Both spectra agree very well.}
\end{figure}

\begin{figure}
\centerline{
\epsfig{figure= 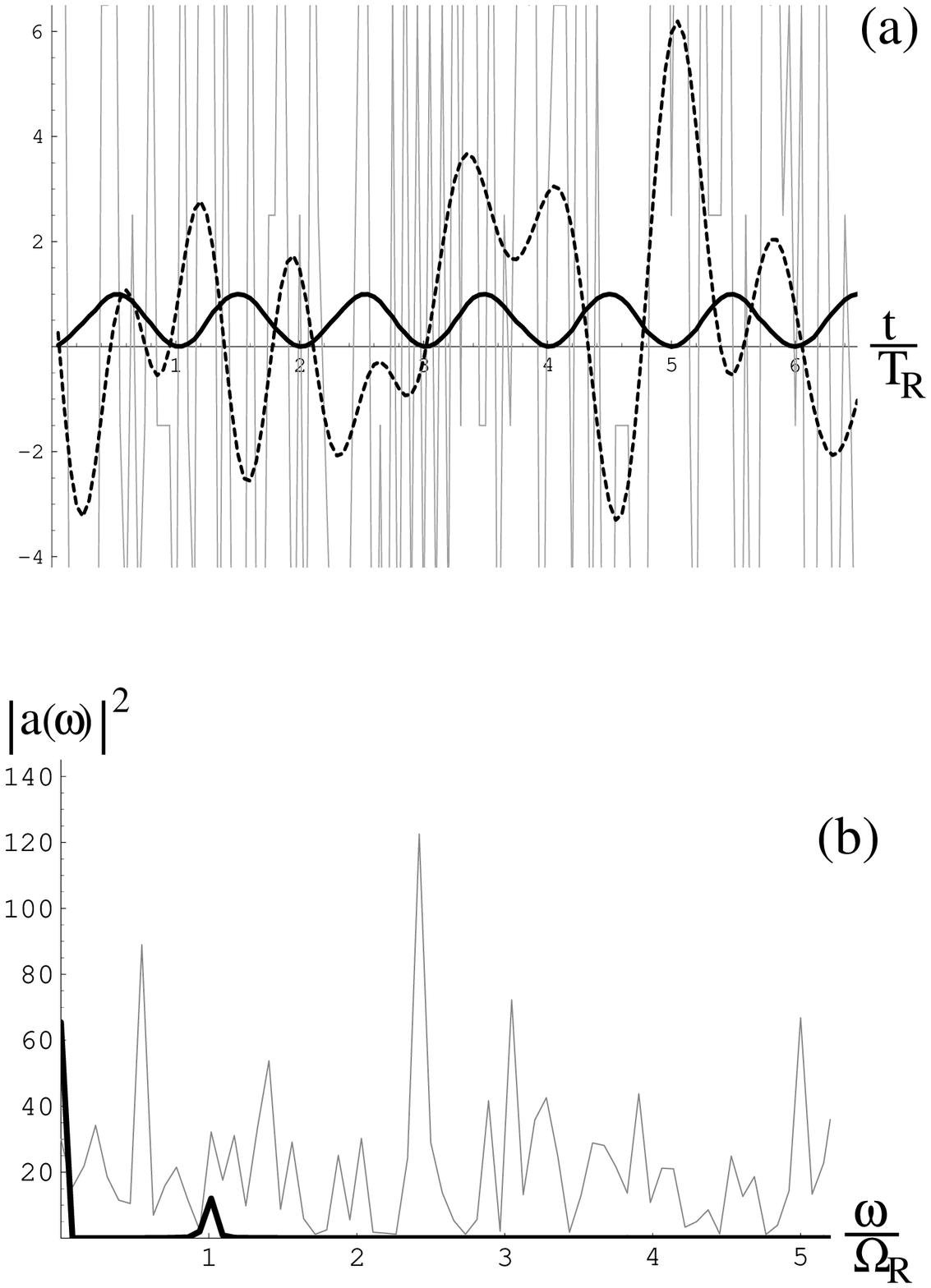, width=0.8\textwidth, angle=0}}
\caption{\label{figure2} Rabi regime with fuzziness $f=62.8$ and $\Delta
p= 0.01\,, p_0=0.5\,,\tau= 0.002T_R\,, N=25$. In Figure (a) the
numerically obtained \ctc{} (black) agrees with the undisturbed Rabi
oscillations. On the other hand $|c_2(t)|^2$  is neither significantly
correlated with the measurement readout (\btc{}, plotted in gray) nor with
the \pbtc{} (dashed) obtained from the readout by noise
reduction. Fig.~(b) displays the power spectrum of $|c_2(t)|^2$
(black) and the power spectrum of the measurement readout (gray). For
graphical reasons the power spectrum of the measurement readout has been
multiplied by a factor of $1/5$. Both spectra are not correlated.}
\end{figure}

\begin{figure}
\centerline{
\epsfig{figure= 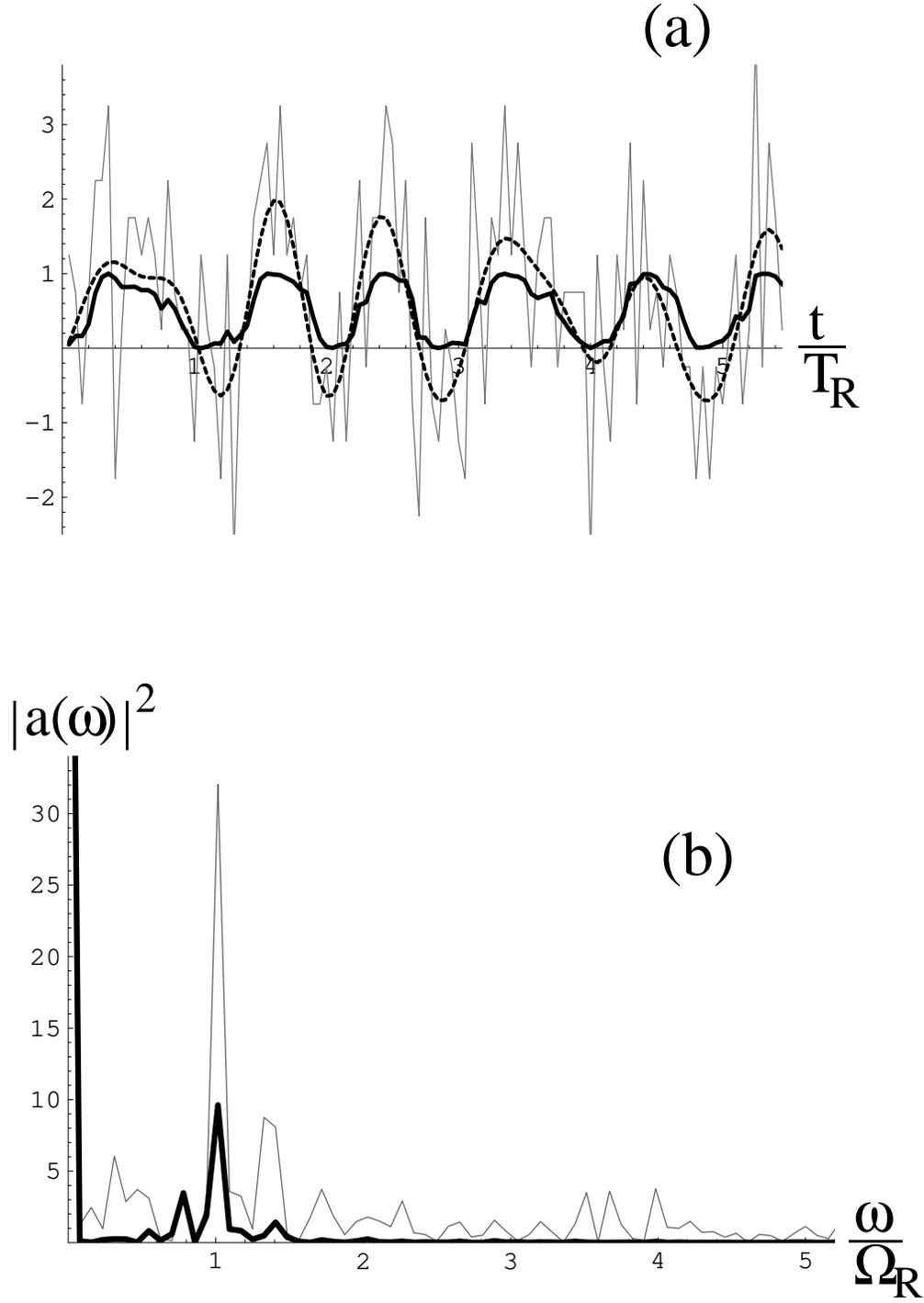, width=0.8\textwidth, angle=0}}
\caption{\label{figure3} Intermediate regime with fuzziness
$f=0.98$ and $\Delta p= 0.08$, $p_0=0.5$, $\tau= 0.002T_R$, $N=25$.
Fig.~(a): There is a high correlation
between the \ctc{} (black) and the measurement readout $G_2(t)$ (gray)
leading to a very good correlation after the noice of the readout has
been reduced (dashed curve). The \ctc{} indicates that the motion of the
state has been disturbed by the measurement. But the Rabi oscillations
are still recognizable. Fig.~(b): The main peak of the power spectrum of the
\ctc{} (black) coincides with the main peak of the power spectrum of the
\btc{}  and indicates approximately
the Rabi frequency $\Omega_R$ of the undisturbed motion.}
\end{figure}

We discuss the three regimes by
considering one typical example for each regime.
The characteristic traits can be read off from
Figs.~\ref{figure1}--~\ref{figure3}. In all three
cases we have chosen $\tau=0.002T_R$ and $N=25$, so that condition
(\ref{Nbound}) is  fulfilled. Figs.~\ref{figure1}a--~\ref{figure3}a show as
function of $t/T_R$ 
\begin{itemize}
\item the $|c_2|^2$-curve of the state motion (black),
\item the \btc{} representing the measurement readout (gray) and 
\item the \pbtc{}
obtained after noise reduction (dashed). 
\end{itemize}
The noise reduction is
sketched in Appendix C. The related
Figs.~\ref{figure1}b--~\ref{figure3}b show the 
respective Fourier spectra  gained by discrete Fourier transform
 of the  $|c_2|^2$-curve (black) and the measurement
readout $G_2$ (grey). The peaks at $k=0$ are
caused by the non-vanishing average value of the curves. In the
three regimes we find characteristically different results:

\subsection{Quantum jump regime}
With fuzziness $f=0.07$ there is a rather 
strong influence of the measurements on the time evolution of the state. The single measurement is still rather
sharp and close to a projection measurement. Accordingly, we observe in
Fig.~\ref{figure1}, that the Rabi oscillations of $|c_2|^2$, which would appear
in the absence of the measurements are completely destroyed. The
$|c_2|^2$-curve (black) shows jumps between extended periods with
values approximately equal to $0$ or $1$. This means that if the state of
the system is close to an eigenstate $\ket{1}$ or $\ket{2}$, it has a
high probability to stay there.  The \btc{} of the
measurement readout (gray) reflects very well the $|c_2|^2$-curve of the state.
The peaks of the $|c_2|^2$-curve pointing upwards and downwards follow the
direction of the corresponding peaks of the \btc{}. The \pbtc{} (dashed)
reflects the quantum jump regime 
even better. In this regime there is a rather good agreement between
the readout and the $|c_2|^2$-curve, but 
there is no characteristic indication of
the undisturbed dynamics. The Fourier spectra show a very good
correlation, as can be seen in Fig.~\ref{figure1}b.
\subsection{ Rabi regime}
The other extreme can be found in the Rabi regime. We have chosen a
high fuzziness $f=63$. The single measurements are
so unsharp and their influence on the motion of the state so
weak, that the Rabi oscillations remain nearly undisturbed. This can be
seen from  the $|c_2|^2$-curve (black)  in Fig.~\ref{figure2}a. In contrast to this 
  the \btc{}  of the measurement readout (grey) is
very fuzzy. The single measurement results lie far away from the
interval $[0,1]$. The processed \btc{} (dashed)  shows also no
correlation, neither in phase nor amplitude  with the $|c_2|^2$-curve. Turning to the
Fourier spectra (Fig.~\ref{figure2}b), we see that the Rabi frequency of the $|c_2|^2$-curve (black)
is represented well. On the other hand, no frequency can be attributed
to the measurement readout (gray curve). In this regime the
measurement readout does not allow a conclusion on the motion of the
state as represented by the $|c_2|^2$-curve.

\subsection{Intermediate regime}
Finally we turn with a fuzziness of $f=0.98$ and $|\Delta p|=0.08$ to a
parameter choice, which leads to representative curves for
the intermediate regime.  According to the Fourier spectrum
(black) in Fig.~\ref{figure3}b the $|c_2|^2$-curve is dominated by one frequency, which differs only little from  the Rabi frequency $T_R$. In this regime the \ctc{},
which would characterize the state motion if no measurement were
performed, is not much modified. The important point is that now the 
\btc{} of the measurement readout (grey) shows a correlation with the \ctc{}
regarding the phase: if the $|c_2|^2$-curve has a maximum, the \btc{} is in the
upper domain and for a minimum correspondingly. Regarding the Fourier
spectrum of the measurement readout (grey curve in Fig.~\ref{figure3}b), there is
still a lot of noise, but a clear peak indicates the frequency of the 
$|c_2|^2$-curve. The measurement readout follows therefore closely the \ctc{} of the
state motion in phase and frequency. This is even more convincing for
the \pbtc{} (dashed curve in Fig.~\ref{figure3}a).                                
\section{Discussion and conclusions}
The normalized state $\ket{\psi(t)} =c_1(t)\ket{1} +
c_2(t)\ket{2}$ of a single two-level system performs oscillations
under the influence of a resonant time dependent driving field. It is
assumed that there is only one realization of this process
available. We ask the question whether it is possible to visualize in
real time the evolution of the system as far as it is given by
$|c_2(t)|^2$. 

This result can clearly not be obtained by means of standard
measurements which result in projections. Generalized measurements,
can be discussed by  means
of positive operator valued measures (POVM). For our goal we restrict
to a particular subclass of unsharp measurements, which have a simple
structure so that an experimental realization should be possible. The
pointer values of this measurements are $+$ and $-$. We perform a sequence
of measurements separated in time leading to the pointer readings 
$+--+-++\ldots$. It can be shown that it is
favorable to transform   a series of $N$ pointer readings to one point of
a measurement readout which is defined as best guess $G_2(t)$ for $|c_2(t)|^2$.

The single unsharp measurement must be weak enough as not to disturb the
evolution of the system too much. On the other hand it must not be too
weak in order to give information about the state of the system. We
determined analytically the domains for the different open parameters
such that this 'intermediate' regime of
measurement can be reached.   The numerical analysis finally
shows that for this intermediate regime the real time visualization
of $|c_2(t)|^2$ can be obtained  to  a
satisfactory amount. The readout $G_2(t)$ reflects rather well the resulting 
$|c_2(t)|^2$, which itself shows only a slight modification of the
undisturbed Rabi oscillations. 

Note that above the Rabi frequency of the unitary development was
assumed to be known. If this is not the case a feed-back mechanism
must be introduced so that after a certain time from the beginning of
the measurements a
visualization can be reached. This project is subject of future
research.
            
\section{Acknowledgment}
The authors thank Peter Marzlin for many stimulating discussions. This
work has been supported by the Optik Zentrum Konstanz.

\begin{appendix}

\section*{Appendix A: How to choose the operations}
In order to detect the dynamics of $|c_2|^2$ we choose measurements
with particular operations corresponding to a particular POVM. We
justify this choice as follows:

Since our measurement scheme contains consecutive measurements, it is
important to consider in which state the system is left after one
measurement. We require the operations, which describe the transformation
of the state up to a phase factor, to be such that the states $\ket{1}$ and $\ket{2}$
are not changed due to the measurement. Hence these states should be
eigenstates of the  operations $M_+$ and $M_-$  in agreement with  (\ref{defM+}) and (\ref{defM-}).

We now address the question why $M_\pm$ should be  positive operators,
thus $u^+_{1,2}\,,u^-_{1,2} > 0$  in
(\ref{defM+}) and (\ref{defM-}) (second requirement of Sect.~\ref{2.1}).   
In the context of quantum feedback control one makes use of the fact
that the operations may be decomposed into 'modulus' and
'phase', in our case:
\begin{equation}
M_\pm=U_\pm |M_\pm|\,,
\end{equation}
where $U_\pm$ are unitary operators and $|M_\pm|:=
\sqrt{M^{\dagger}_\pm M_\pm}$. While the 'modulus' $|M_\pm|$, because of its
relation to the effects $E_\pm=|M_\pm|^2$, is connected to the
acquisition of information, the unitary operators $U_\pm$ represent rather an additional
Hamiltonian evolution, not leading to any increase of information
about the system (compare Wiseman \cite{Wiseman} and Doherty et al.~\cite{5}).
If thus $U_\pm$ cannot be used as a feedback to compensate
the influence of $|M_\pm|$, it only adds to the disturbance by the 
measurement and should be set $U_\pm=1$ (a global phase factor
$\exp(i\varphi_\pm)$ is always neglected).

That the additional evolution represented by $U_\pm$ is indeed obstructive in
our case can be made clear as follows.  
To be diagonalizable the $M_\pm$ have to
be normal, i.e., $[M^\dagger_{\pm}, M_\pm]=0$ which amounts to
$[|M_\pm|^2,U_\pm]=0$. Hence $U_\pm$ and $|M_\pm|$ have a simultaneous basis of
eigenvectors, which has to be $\ket{1}$ and $\ket{2}$, because $M_\pm$ is
required to be diagonal in that basis. Now if $U_\pm$ is diagonal with
respect to $\ket{1}$ and $\ket{2}$, it can be
written as:
\begin{equation}
U_\pm =\exp(-\frac{\mbox{i}}{2}\sigma_z\theta_\pm)     
\end{equation}
for some angle $\theta$ and $\sigma_z$ as Pauli operator. 
In order to illustrate the corresponding different influences on 
the state motion,
we refer to the Bloch sphere. 
The resonant Rabi oscillations which our system performs in the absence of
measurements can be represented by a rotation of the normalized Bloch
vector  in the y-z-plane (given that the initial state
lies in this plane) . 
The measurement induces a
change of the state given by $M_\pm$. The contribution due to
$|M_\pm|$ leads to a jump of the Bloch vector towards one of the
'poles' on the Bloch sphere at $(0,0,1)$ and at $(0,0,-1)$. The Bloch
vector thereby does not leave the plane spanned by the state before the
measurement and the z-axis. $U_\pm$ on the other hand forces the Bloch
vector to leave this plane. As a consequence $U_\pm$ cannot compensate the
change due to $|M_\pm|$ and  the influence of $|M_\pm|$ is
inevitable. Beyond that, a unitary evolution $U_\pm\not=1$ disturbs the
original Rabi oscillations
heavily in a succession of measurements by leading the Bloch vector
out of the y-z-plane. Therefore  we have  to  choose $U_\pm=1$ in order to
disturb the Rabi oscillations by the action of the operations $M_\pm$ as 
little as possible. In an experimental situation with nontrivial unitary 
transformations $U_{\pm}$ the state change due to $U_{\pm}$ could be 
compensated by means of feedback (compare Wiseman \cite{Wiseman}).

\section*{Appendix B: Upper limit for $N$}
In this Appendix we describe under which circumstances the concept of
the best guess as introduced in Sect.~\ref{3} can also be used  for
N-series with dynamical evolution due to the Hamiltonian in (\ref{H})
between the single measurements.

Consider a N-series of unsharp measurements which are  separated by the
unitary evolution according to (\ref{H}). In the interaction picture the
state after the first N-series thus reads:
\begin{equation}
\label{totev}
\prod_{k=1}^N\,\left(M_k \exp\left(-\frac{i}{\hbar}H_I\tau\right)\right)
\ket{\psi(0)}
\end{equation} 
with $M_k = M_+,M_-$ and $H_I:=\hbar\Omega_R(\ket{2}\bra{1}
+\ket{1}\bra{2})/2$. We  have to discuss for this new
situation again the concept of a best
guess. If we may commutate unitary evolution
and operations representing the change due to the measurement, so that 
\begin{equation}
\label{sep}
\ket{\psi(\Delta t)}\approx  \exp\left(-\frac{i}{\hbar}H_I\Delta
t\right) \tilde{M}(N_+,N)\ket{\psi(0)}
\end{equation} 
with 
\begin{equation}
\label{Mtilde}
\tilde{M}(N_+,N) := \sqrt{{N \choose{N_+}}}M(N_+,N)\,,
\end{equation}
we are back to the situation described in  Section \ref{3.2} and the best
guess $G_2(t_m)$ of (\ref{b2}) may be taken as a good candidate to
approximate  $|c_2(t_m)|^2$.
The underlying commutators are
\begin{equation}
\left[M_\pm\,,\,
\exp\left(-\frac{i}{\hbar}H_I\tau\right)\right]=\sigma_y(u_1^{\pm}-u_2^\pm)
\sin(\frac{\pi\tau}{T_R})\,, 
\end{equation}
where $\sigma_y$ represents the Pauli operator. Although the
commutators are small for $\tau/T_R\ll 1$, their contribution 
accumulates and is no longer negligible if the
number $N$ of measurements within the N-series is too large.  

A lengthy calculation shows that for $N\tau/T_R\ll 1$ the term
(\ref{totev}) may be 
approximately written in the form
(\ref{sep}), if 
\begin{equation}
\label{Nbound}
(N-1)^2 \ll \frac{\max\{u_1^+, u_2^-\}}{
2\max\{|u_2^+-u_1^+|,|u_2^--u_1^-|\}}\frac{T_R}{\pi\tau}\,.
\end{equation}
Here it was assumed that the matrix elements of the operation $M_+$
fulfill $u_1^+>u_2^+$. Otherwise the indices $1$ and $2$ in
(\ref{Nbound})
have to be permutated.

\section*{Appendix C: Simulation procedure and noise reduction}
In this appendix the procedure is described we employed to simulate
the measurements of Rabi oscillations on the computer.   
It is denoted here in form of four instructions:\\
1) Evolve initial state   unitarily according to the
undisturbed dynamics of the system; $\ketpsi{0}\rightarrow
\ketpsi{\tau}=\exp\{-\frac{\mbox{i}}{\hbar} H_I\tau\}\ketpsi{0}\,.$\\
2) Simulate single unsharp measurement in two steps. First generate a
random outcome by a Monte Carlo method according to the probabilities
$p_\pm=\ex{E_\pm}_{\psi(\tau)}$. Then 
change the state depending on the outcome $\ket{\psi(\tau)}\rightarrow
M_\pm\ket{\psi(\tau)}$.\\   
3) N-series: 
Carry out instructions 1) and 2) $N$ times, every time  
replacing the  initial
state by the state obtained after the previous unsharp measurement. 
After the N-series record $|c_2(t_1=N\tau)|^2$ of the current state and the
measurement readout, i.e., the best guess $G_2(t_1)$ for
$|c_2(t_1)|^2$. 
The best guess is calculated from the number of $+$-results
(c.f. equation (\ref{b2})).\\
4) Simulate $M$ times an N-series of unsharp measurements, always
inserting as initial state the state obtained from the previous N-series.
The total record of the simulation will then consist of a sequence of  readouts
$G_2(t_m)$  and squared components of the state
$|c(t_m)|^2$ with $t_m= m\Delta t\,, m=1...M$.     

The readouts from the simulated measurements can now be processed to
eliminate noise. We used the following method.\\
The sequence $(G_2(t_m))$ is expressed by means of discrete
Fourier transform as
$G_2(t_m)= \sum_{l=0}^{M-1}\,a(\omega_l) \exp(i\omega_l
t_m)$, with $\omega_l= 2\pi l/T$. 
From the power-spectrum ($|a(\omega_l)|^2$) the main peak (if present) is
identified with the modified Rabi frequency and the noise is estimated.
 
The noise is reduced by transforming the Fourier coefficients
employing a Wiener filter $\phi_l$ \cite{13}:
$a(\omega_l)\rightarrow a(\omega_l)\phi_l$.  
Further noise reduction is achieved by omitting higher frequencies in
the Fourier series which cannot be related to the original Rabi
oscillation. We truncated the Fourier series after the $2l$-th term,
where $\omega_l$ is the frequency of the main peak of the
power spectrum. 

The resulting filtered and truncated Fourier series
leads to a new sequence of best guesses $(G_2(t_m))$ in the text
referred to as 'processed measurement readout' or
'processed $G_2$-curve'.

\end{appendix}

\end{document}